\documentclass[apj]{emulateapj}

\begin{document}

\title{The Cosmic Ray Intensity Near the Archean Earth}

\author{O. Cohen\altaffilmark{1}, J.J. Drake\altaffilmark{1}, J. K\'ota\altaffilmark{2}}

\altaffiltext{1}{Harvard-Smithsonian Center for Astrophysics, 60 Garden St. Cambridge, MA 02138, USA}
\altaffiltext{2}{Lunar and Planetary Laboratory, University of Arizona, Tucson, AZ 85721-0092, USA}

\begin{abstract}

We employ three-dimensional state of the art magnetohydrodynamic models of the early solar wind and heliosphere and a two-dimensional model for cosmic ray transport to investigate the cosmic ray spectrum and flux near the Archean Earth.   We assess how sensitive the cosmic ray spectrum is to changes in the sunspot placement and magnetic field strength, the large scale dipole magnetic field strength, the wind ram pressure, and the Sun's rotation period.   Overall, our results confirm earlier work that suggested the Archean Earth would have experienced a greatly reduced cosmic ray flux than is the case today. 
The cosmic ray reduction for the early Sun is mainly due to the shorter solar rotation period and tighter winding of the Parker spiral, and to the different surface distribution of the more active solar magnetic field.  These effects lead to a global reduction of the cosmic ray flux at 1~AU by up to two orders of magnitude or more.  Variations in the sunspot magnetic field have more effect on the flux than variations in the dipole field component.   The wind ram pressure affects the cosmic ray flux through its influence on the size of the heliosphere via the pressure balance with the ambient interstellar medium.   Variations in the interstellar medium pressure experienced by the solar system in orbit through Galaxy could lead to order of magnitude changes in the cosmic ray flux at Earth on timescales of a few million years.
\end{abstract}



\section{INTRODUCTION}
\label{sec:Intro}

During the Archean eon ($\sim3.8-2.5$Gya) when life on Earth is thought to have first emerged, the Sun's rotation period was 6-15 days, in contrast to its current 27 day period \citep{Skumanich72}.  A rotation-activity relation is well known from observations of young cool stars, in which the faster the rotation the higher the magnetic activity and related UV-X-ray luminosity  \citep[e.g.][]{Pallavicini81,gudel07}.  Young active stars are also observed to have magnetic spots concentrated at high latitudes \cite[i.e. polar regions,][]{Strassmeier01}, in contrast to the low-latitude spots that characterize the magnetic activity of the present day Sun. Since the topology of the solar surface magnetic field governs the topology of the interplanetary magnetic field \citep[IMF,][]{Parker58c}, and the topology of the solar wind \citep[see summary by][]{McComas07}, the interplanetary environment of the young Sun would have been very different from the one at present.   \citet{Parker58b} showed that cosmic ray (CR) transport within the solar system, and the flux of energetic particles that reaches earth, depends on the IMF.  Consequently, even in the absence of temporal and spatial variations in Galactic CR production, we can surmise that  the Archean earth likely experienced a quite different CR exposure to that of the present.

As an ionizing radiation source, CRs could have played a significant role in the origin and evolution of life on Earth.   Their potential influence includes 
inducing a varied and complex organic chemistry, including possible production of nucleotides
\citep[e.g.][]{Court06,Simakov02}, causing cellular radiation damage and mutation through direct and indirect processes \citep[e.g.][]{Nelson03,Dartnell11} 
aiding atmospheric lightening initiation \citep[e.g.][]{Gurevich99}, and (perhaps more controversially) affecting climate through production of cloud condensation nuclei \citep[e.g.][]{Svensmark97,Shaviv03,Wallmann04,Shaviv05,Kirkby11}.  See also \citet{Medvedev07} for an overview.  Understanding the way in which solar magnetic activity has influence the CR exposure of Earth through time is then of interest from a variety of both physical and biological perspectives.

The modulation of galactic CRs in the heliosphere is described by the diffusive transport equation of \cite{Parker65}. The omnidirectional phase space density $f(\mathbf{r},p,t)$ at position $\mathbf{r}$, momentum, $p$, and time, $t$ obeys:
\begin{equation}
\frac{\partial f}{\partial t} = \nabla (\kappa \nabla f)
- (\mathbf{V} + \mathbf{V}_d ) {\nabla f}
+ {\nabla \mathbf{V} \over 3} \frac{\partial f }{\partial \ln{p}},
\end{equation}
where $\kappa$ stands the anisotropic diffusion tensor with different components parallel and perpendicular to the Parker-spiral magnetic field ($\kappa _{\parallel}$, and $\kappa _{\perp}$, respectively). $\mathbf{V}$ is the convective solar wind velocity,  while $\mathbf{V}_d$ describes large-scale gradient and curvature drifts due the regular gyration of CR particles in the non-uniform field. The last term accounts for the energy change due to a deceleration in the expanding wind and acceleration at the Termination Shock (TS).



\citet{Svensmark06} calculated the CR spectrum at 1~AU back in time using scaling laws for the solar wind and mass loss rate as a function of time, and a scaling law for the average solar surface magnetic field as a function of rotation rate. 
Due to the lack of direct observations of solar-like winds of cool stars, it is not clear to what extent the scaling laws used by \citet{Svensmark06} are valid.  \citet{Wood02}, for example, caution against extrapolating their secular wind mass loss relation that was used in the \citet{Svensmark06} study to significantly higher magnetic activity levels.  

The \citet{Wood02} relation is based on a correlation between X-ray and wind fluxes for a handful of stars, and observed secular decline of stellar X-ray luminosity.
\citet{Cohen11b} argued that stellar mass loss rate should not be scaled with stellar X-ray luminosity, since they are determined by different components of the stellar magnetic flux (X-rays being associated with closed flux, and mass loss with open flux).  In short, wind mass loss rates for the early Sun are very uncertain.
\cite{Sterenborg11} calculated a series of scenarios for the IMF and the solar wind structure for the young Sun.  Using a magnetohydrodynamic wind model and a surface magnetic field consistent with observations of young active stars, they found a wind mass loss rate not more than 10 times that of the current Sun.  This result is consistent with the theoretical study of \cite{Holzwarth07}.  

Here, we extend the work done by \cite{Sterenborg11}, and build on their wind and IMF model solutions to investigate the CR spectrum near the Archean Earth using a two-dimensional model for transport of CR in the heliosphere.  In particular, we focus on the effect of latitudinal variations of solar active regions and the enhanced solar rotation.  We do not assume anything about the effect of the TS and the heliopause, or the location of the solar system within the Galaxy.  In other words, we assume the same incoming CR distribution for each possible solar magnetic field topology.

In Section~\ref{sec:Model}, we describe our experimental approach and the assumptions being made in both modeling the solar wind and the IMF of the young Sun, and the calculation of CR transport. We describe the results in Section~\ref{sec:Results} and discuss their meaning in Section~\ref{sec:DiscussionConclusion}. 



\section{MODEL DESCRIPTION}
\label{sec:Model}

Our calculation of the CR flux near the Archean Earth involves building a model of the solar wind and the IMF for the young Sun, which was done by 
\citet{Sterenborg11}, 
and the calculation of the CR transport from the edge of the solar system to the Earth based on the output from the solar wind model. The two steps are as follows.

\subsection{Solar Wind and the Interplanetary Magnetic Field}
\label{SWIMF}

The transport of CR depends mostly on the topology of the magnetic field in the heliosphere (i.e., the IMF). The IMF is governed by the surface distribution of the solar magnetic field, which is carried away by the solar wind, and wound up by the solar rotation.  \cite{Sterenborg11} have used the BATS-R-US Magnetohydrodynamics (MHD) code \citep{powell99, toth05, cohen07} to calculate the solar wind and the IMF conditions for the young Sun, based on a set of scenarios for the young Sun's surface field topology, as well as for different rotation rates. For each solution, a synoptic map of the solar surface field (magnetogram) was used to calculate a steady-state, three-dimensional solution for the solar corona and the solar wind. 

\begin{figure*}[h!]
\centering
\includegraphics[width=6.5in]{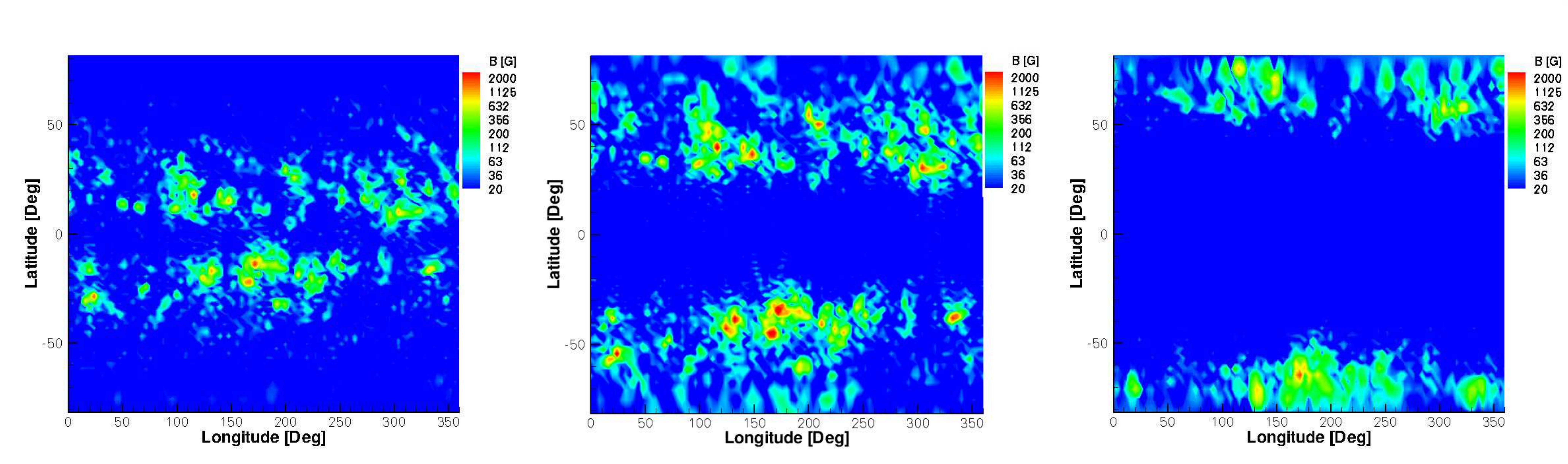}\\
\includegraphics[width=6.in]{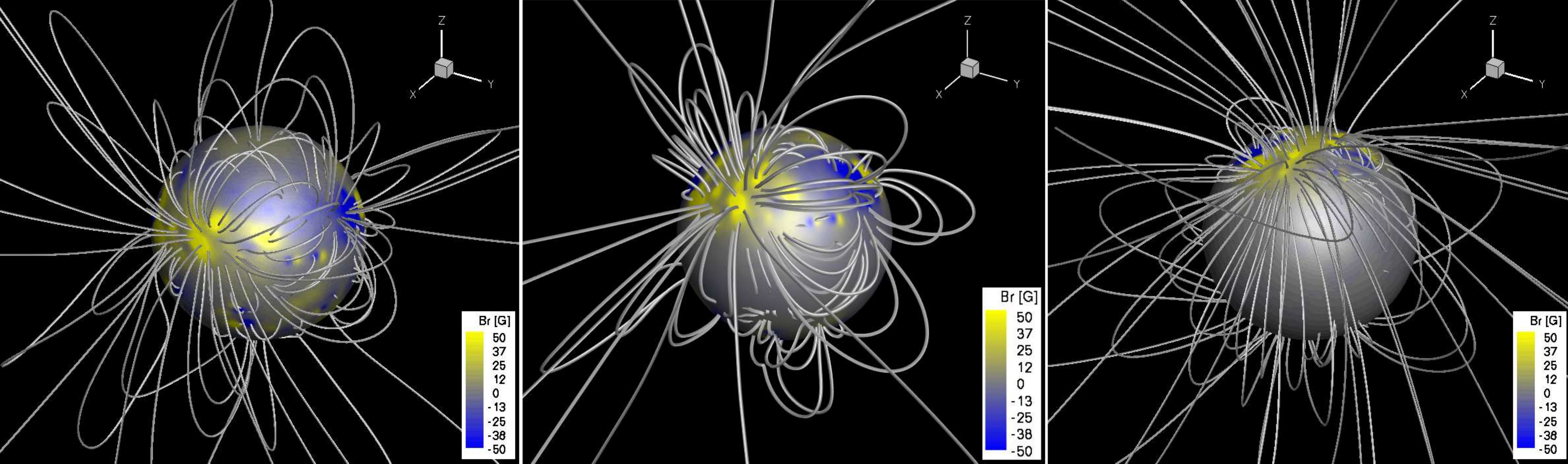}
\caption{The observed surface distribution of the photospheric radial magnetic field during 
solar maximum is shown on the top left, while the manipulated magnetogram with the spots shifted by 
30 and 60 degrees toward the poles is shown on the top middle and right panels, respectvely.  The three-dimensional magnetic field topologies corresponding to these three surface field distributions (displayed on each sphere) are shown on the bottom.}
\label{fig:f1}
\end{figure*}

For the young Sun, a magnetogram of the current Sun during solar maximum was used as the basis for a reference case. The magnetogram was manipulated and split into two components---a weak dipole component, denoted by a Dipole Factor (DF), and a strong spot components, denoted by a Spot Factor (SF).  In order to resemble observations of young active stars, the active regions in the map were shifted by 30 and 60 degrees towards the poles, so that they appeared at mid-high and high latitudes. The case of 60 degrees shift should represent the solar magnetic field at the begining of the Archean , while the 30 degrees shift should represent the field towards the end of that period \citep[e.g.,][]{TonerGray88}. Figure~\ref{fig:f1} shows the magnetograms and the three-dimensional magnetic field structure for the reference case of the current Sun and for the reference young Sun cases with the spots manipulated and shifted poleward.  Other test cases involved scaling the magnitude of either magnetogram components (DF or SF) by a factor of 10.

Here, we use these solutions to calculate the transport of CR from the edge of the solar system to 1AU.  Since we are interested in the average state of the young Sun's IMF, longitudinal variations are not important.  This offers some computational expedience.  We found the wind solutions are fully developed and super-Alfv\'enic at a radial distance of $20R_\odot$, such that a much more computationally expensive fully-detailed calculation out to 1AU is not necessary for understanding the longitudinally-averaged wind at at Earth.  We therefore extract the physical parameters from the model at $20R_\odot$, averaged over longitude, and extrapolate the model to the TS.
We have computed models for rotation periods of 2, 4.6, 10, and 26 days to probe the effects of rapid rotation and winding of the Parker spiral.  

The beginning of the Archean, about 3.8~Gya,  corresponds to solar ages comparable to open clusters such as M37 (550~Myr) and the Hyades (625~Myr), when the solar rotation period would have been about 4-6 days \citep[e.g.][]{Meibom11}, while throughout the Archean, the solar rotation period was increasing to 10-15 days \citep{telleschi05}.

\subsection{Cosmic Ray Transport}
\label{CRtransport}

The modulation of CR is simulated in an updated version of the quasi-steady-state,
two dimensional numerical model by \cite{Jokipii93}. An average uniform radial solar wind with the value of the average wind speed is taken from each of the different young Sun models described in Sect.~\ref{SWIMF}.
The TS is placed at 90~AU for the current Sun, and at distances scaling with the square root of the average ram-pressure for other cases (95~AU, 110~AU, and 120~AU for polar spots, enhanced DF, and enhanced SF, respectively).  We perform additional calculation of all cases with the TS remain fixed at 90~AU for reference.

We neglect azimuthal gradients, and use the azimuthally averaged values of the magnetic field for the diffusion coefficients, $\kappa _{\parallel}$ and $\kappa _{\perp}$, as well for the drift velocity, $\mathbf{V}_d$. Though this is not a strictly valid assumption, it serves as a good approximation for our purposes to estimate the omnidirectional CR flux.  For the diffusion coefficients, we adopt the simplest scaling
law, assuming that both  $\kappa _{\parallel}$ and $\kappa _{\perp}$ are proportional to the particle speed times the gyroradius, i.e.\ they are scaled inversely with the large-scale field, $B$. We take  $\kappa _{\parallel} = 18.1 \beta (P/B)\; AU^2/day$, and $\kappa _{\perp} = 0.05 \kappa _{\parallel}$. Here, $\beta = v/c$ is the particle speed as a fraction of the speed of light, $P= pc/Ze$ is the particle rigidity, which is essentially the momentum per unit charge, and $B$ is the magnetic field strength in units of nT.  We also add a transverse field component impeding easy penetration through the weak spiral field through the polar regions according to 
\cite{JokipiiKota89}.

The omnidirectional density, $f$, relates to the particle flux, $J_T$, through $J_T=4 \pi f p^2$, and $J_T$ is set at the outer boundary to the interstellar spectrum given by \cite{WebberLockwood01} as:
\begin{equation}
J_T= \frac{21.1 T^{-2.8}}{1.+5.85 T^{-1.22} + 1.18 T^{-2.54}}.
\end{equation}
with $T$ being the particle kinetic energy.

One omission of the present work is the treatment of CR transport through the heliopause and the TS itself, since we are interested in the CR transport {\it within} the solar system.


\section{RESULTS}
\label{sec:Results}

\begin{figure*}[h!]
\centering
\includegraphics[width=7.in]{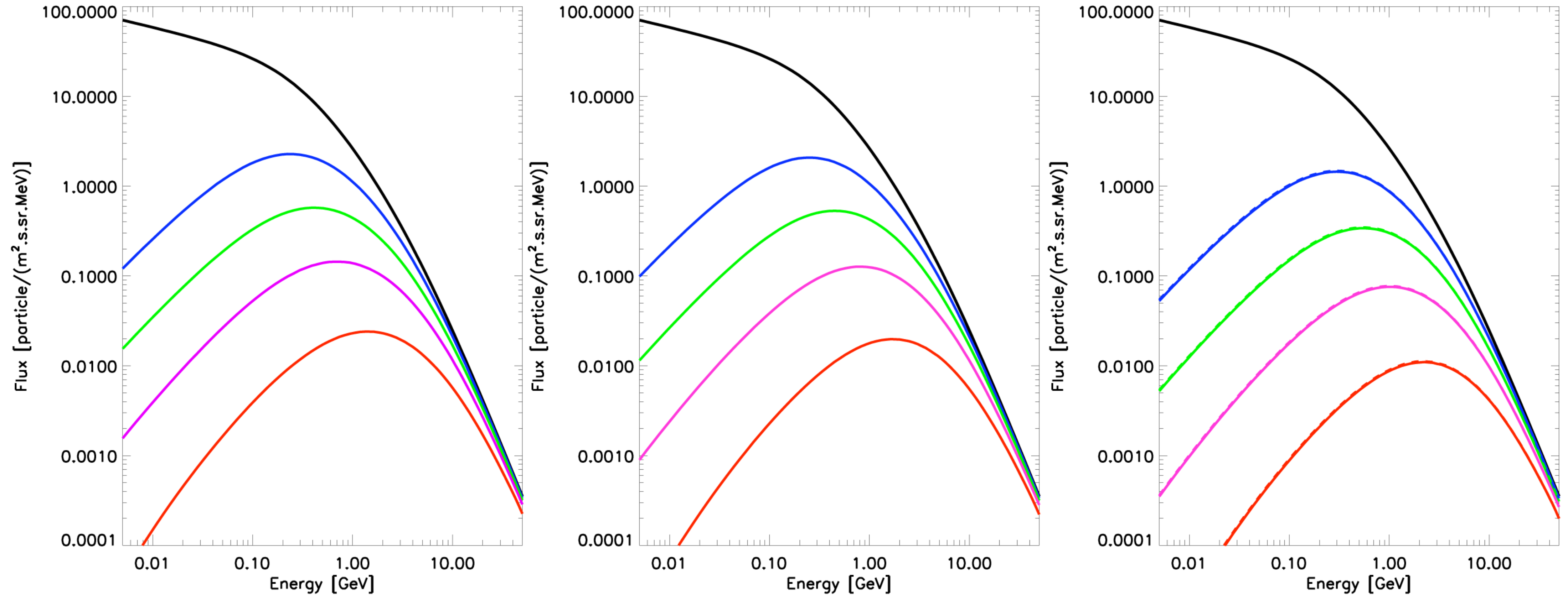}
\caption{CR energy spectrum for solar rotations of 26d (current rotation, blue), 10d (green), 4.6d (magenta), and 2d (red). The local ISM spectrum is shown in black.  Plots are for the current Sun (left), spots shifted towards the pole by 30 degrees (middle), and spots shifted towards the pole by 60 degrees (right). Solid lines represent the spectrum with the TS scaled with the solar wind dynamic pressure, while dashed lines represent the spectrum for the TS fixed at 90~AU.}
\label{fig:f2}
\end{figure*}

The results for the CR energy spectrum at 1~AU for the simulated test cases are shown in Figure~\ref{fig:f2}. Each plot shows the simulated CR spectrum for the different solar rotation periods. The plots also show the effect of variations in the location of the TS for the cases representing the young Sun. The result of the spectrum for the current Sun with the current rotation period are consistent with those presented in Figure 1 in \cite{Svensmark06} (presenting both model calculation and data taken from \cite{McDonald01}), with the flux peak around $1-2\;[particles\;m^{-2} s^{-1} sr^{-1} MeV^{-1}]$ located around $300-500\;MeV$. 

\begin{figure*}[h!]
\centering
\includegraphics[width=6.in]{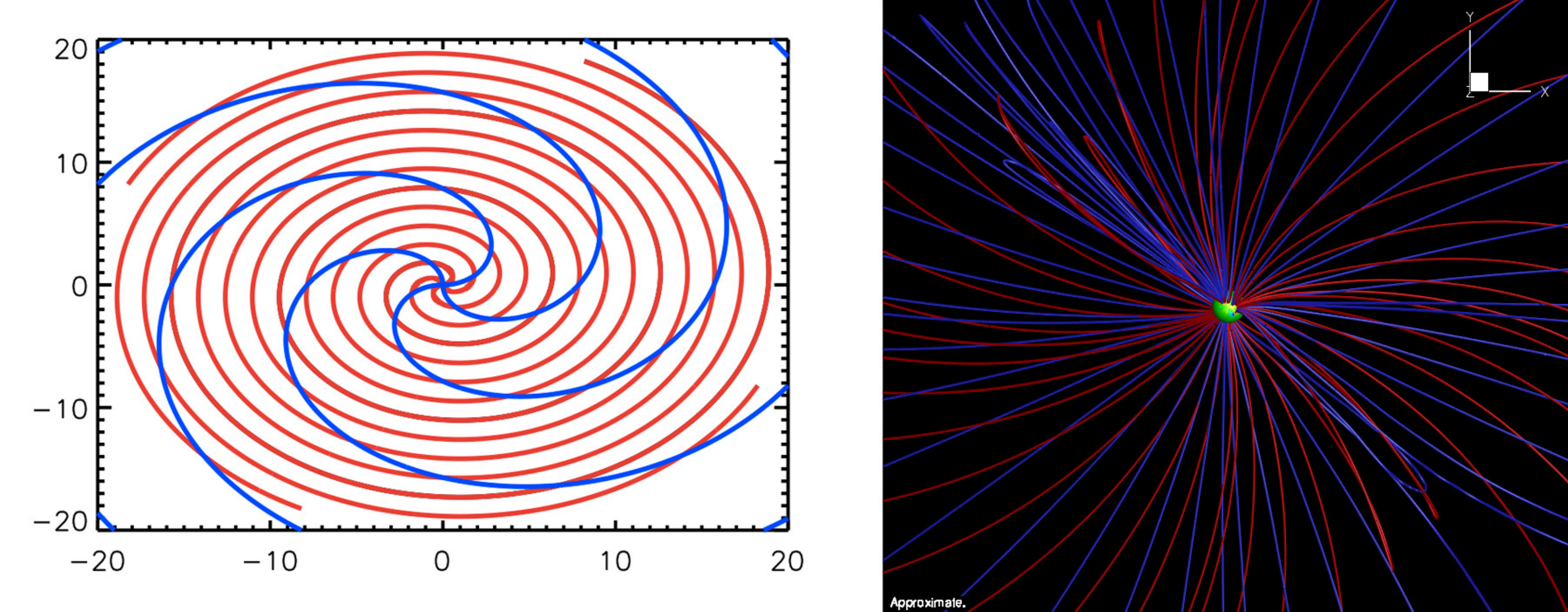}
\caption{Left: conceptual display of different spirals with slow (blue) and fast (red) rotation rates as a function of distance. Right: the simulated equatorial magnetic field near the Sun (up to $24R_\odot$) for the case with solar rotation period of 4.6 days (red) and 26 days (blue) displayed from the top. The difference in tangling increases with distance from the Sun.}
\label{fig:f3}
\end{figure*}

The most dramatic change, which is clearly seen in all panels, is the change in the spectrum due to the enhanced solar rotation. The faster rotation winds up the field carried by the wind and increases the azimuthal component of the IMF.  The enhanced and compressed Parker spiral prevents CR penetration to 1~AU much more efficiently than the Parker spiral of the current Sun.  Even for the relatively low activity of the present day Sun, the enhanced rotation reduces the flux at the peak of the spectrum by more than an order of magnitude. It also shifts the spectrum peak by a factor of 2 towards higher energies from about 500~MeV to 1~GeV. Figure~\ref{fig:f3} shows the equatorial magnetic field for the cases with solar rotation periods of 4.6 and 26 days, along with a conceptual display of spirals with different rotation rates. While the difference in the field tangling is notable in the simulation box extended up to $24R_\odot$, it becomes larger at greater heliospheric distances.  

The other effect that is clearly seen from the results is that of the sunspot distribution, and of the magnitude of the different field components.  By only moving the spots by 60 degrees towards the pole, a slight---30-50\%\ or so---but global decrease in the CR flux is obtained. This change is due to the increase in the polar field as the spots occupy high-latitude regions.  Since it is easier for CR to penetrate through polar regions, where the field is more radial and relatively weaker, an increase in the polar field leads to a reduction of the CR flux.  The shift by 30 degrees does not seem to change the CR flux at all, probably since the shifted spots do not interact and modify the polar field as in the 60 degree shift case.

\begin{figure*}[h!]
\centering
\includegraphics[width=6.in]{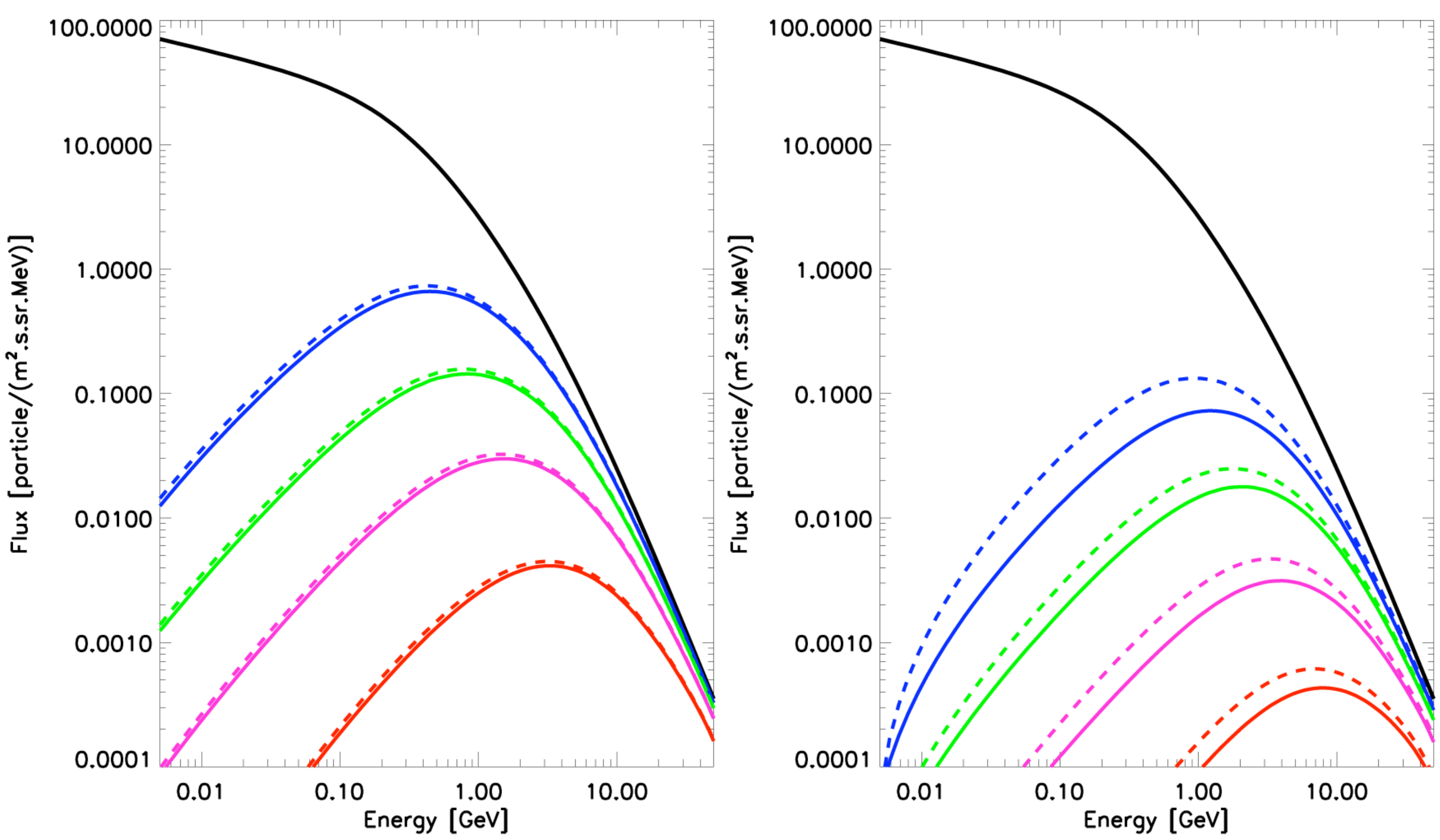}
\caption{Same as the right panel of Figure~\ref{fig:f2}, but with the dipole component enhanced by a factor of 10 (left), and the spot component enhanced by a factor of 10 (right).}
\label{fig:f4}
\end{figure*}

Figure~\ref{fig:f4} shows the results for the cases with enhanced dipole and spot components of the magnetic field. When the dipole component of the reference young Sun polar spot model is enhanced by a factor of 10 (a polar field of about 50~G), the spectra for the slow to fast rotation periods are suppressed by factors of about 2 and 4-5, respectively, with only very modest changes in the spectral peaks. For the case where the polar spot component is enhanced by a factor of 10 (field strengths on small spatial scales of few kG), both slow and fast rotation spectra are decreased much more dramatically---by almost an order of magnitude. In fact, all CR with energies less than 100~MeV are completely eliminated from the spectrum for the cases with 4.6 and 2 days rotation period.

The location of the TS does not have a large influence on the spectrum, except for the case of enhanced polar dipole/spots where it is moved from 90~AU (the current Sun) to 120~AU.  The TS being further out causes the CR to travel longer distances from the edge of the solar system to 1~AU.  This results in a larger CR loss on the way and an overall reduction in the CR flux that reaches the inner solar system.  The change in the CR spectrum for the other models is smaller partly because their wind solutions are more similar in terms of wind momentum to that of the present day active Sun case and the TS therefore not so far out.  For the polar spots model, CR coming from high latitudes are also attenuated more because of the significantly enhanced polar field. 


\section{DISCUSSION AND CONCLUSIONS}
\label{sec:DiscussionConclusion}

Our simulations demonstrate how the different interplanetary environment during the early Archean would have resulted in a quite different CR spectrum and greatly reduced flux near the Earth than we experience today.   What we have done here is to isolate the different aspects of the magnetic activity of the Sun to determine to what extent each affects the resulting CR spectrum.  
The CR reduction we find is mainly due to the enhanced solar rotation and the different surface distribution of the more active solar magnetic field.  These effects lead to a global reduction of CR flux at 1~AU by up to two orders of magnitude or more, and to elimination of the low-energy ($<100MeV$) flux in the most extreme 2 day rotation period scenario.   The results are more sensitive to the assumed strength of polar spots than to the large-scale dipole field.   

\subsubsection{Comparison with previous work}

\citet{Shaviv03} estimated that a stronger solar wind from a more active young Sun would effect a reduction in the CR flux reaching Earth that could have influenced the climate.   The results presented here illustrate that a stronger wind, giving a larger heliosphere and TS shock distance is not the main effect; instead the more rapid solar rotation that winds the Parker spiral is the dominant screening mechanism. 
 
It is useful to compare our results with those presented in Figure~1 of \citet[][]{Svensmark06}; we have adopted the same units for displaying the cosmic ray energy spectrum as that study.  We generally confirm their findings, though with some differences.  
In their curve for 3.8 billion years ago, almost the entire CR energy spectrum is eliminated except for energies above about 10~GeV where the flux is reduced one hundredfold.  This is similar to our last case in Figure~\ref{fig:f4} with a strong polar (spot) field and a rotation period of 2 days---more extreme, at least in terms of rotation, than we believe the Sun at that age would have been.  As noted in Sect.~\ref{sec:Model}, at the begining of the Archean (the Eoarchean era), the solar rotation period would have been 4-6 days, which seems somewhat slower than adopted by \citet{Svensmark06}.   

In \citet{Svensmark06}, the effect of the magnetic field distribution was treated by simply scaling the average surface field with the rotation rate, with a resulting field strength as a function of time, $t$, of $B(t)=B_0t^{-0.6}$, with $t$ in units of 4.6~Gyr.  At 1 Gyr, this formula predicts a field 2.5 times that of today---a fairly modest difference.  Our simulations enable us to separate the effects of rotation rate, field distribution, field strength, and the location of the TS.  Therefore, they provide intermediate solutions as well. In some of our scenarios, the CR flux does not become negligible, especially in the lower energy range. This contrasts with the results of \citet{Svensmark06}.

\subsection{Surface Field of the Archean Sun}

The sensitivity of our results to the assumed polar field strength---comparing the panels in Figure~\ref{fig:f2} and \ref{fig:f4}---raises the question of what polar spot field strength is appropriate for the young Sun?

The polar regions of stars are difficult to probe in detail with Doppler imaging techniques.  These methods employ rotationally-modulated wavelength shifts in Zeeman signatures to map the surface field structure \citep{Donati09}.   Polar regions exhibit only small projected line-of-sight velocity changes through a rotation cycle (or none at all at the pole itself), and consequently the resolution of the mapping technique is much more limited.  In particular, a polar region with mixed polarity field, as characterizes our young Sun reference model, will unlikely register as a strongly magnetic region because the opposite polarity Zeeman signatures will cancel out.  Moreover, the dark polar spots commonly seen on active stars contribute little to the full-disk spectrum and can be essentially invisible to Zeeman-Doppler imaging \citep[e.g.][]{Marsden06a}.  Indeed,  \citet{Marsden06a}, \citet{Marsden06b} and \citet{Jeffers08} found polar regions with field strengths of a few hundred G circling a dark polar spots within which no significant field was detected on the 30--50~Myr old, solar analogs HD~171488 ($P\sim1.3$d) and HR~1817 ($P\sim 1.0$d).    At somewhat later spectral types that might not be as relevant to the solar case, the well-studied very active zero-age main-sequence K0 dwarf AB~Doradus (age $\sim 75$~Myr; \citealt[e.g.][]{Janson07})  shows some mixed polarity regions at $50\deg$ latitudes with field strengths up to 500~G \citep[e.g.][]{Hussain07,cohen10b}.  

Larger scale fields are more easily resolved.  \citet{Petit08} found a mean large scale field of $51\pm 6$G for the solar analog HD~190771, which has a rotation period of 8.8d.  This is an order of magnitude larger than that of the Sun, with one third the rotation period.  

It seems likely, then, that both the large scale dipolar and smaller scale spot fields on the young Eoarchean Sun were significantly stronger than seen on the Sun of today.  The model with a 4 day rotation period and spot and fields enhanced by a factor of 10 is then probably not too far from the truth. While this model produces peak field amplitudes of more than a kG, the stellar observations are limited in spatial resolution and would tend to smear and average out field with fine structure. As the Sun aged and was rotationally braked by the solar wind through the Archean, the spot and large scale field strengths would have declined and shifted towards lower latitudes.


\subsubsection{Broader Relevance}

The relevance of CR for modulation of the Earth's climate---first raised as a possibility by \citet{Ney59}---has not yet been placed on a firm footing.  Early evidence suggesting a CR-climate connection was summarised by \citet{Shaviv03}.  The general mechanism is thought to be the seeding of cloud condensation nuclei by CR ionization events in the lower atmosphere that results in a higher albedo and cooler average temperatures.  While several studies have refuted the connection \citep[e.g.][and references therein]{Erlykin09,Pierce09}, recent experimental support for the basic mechanism has been presented by \citet{Kirkby11}.  

Overall, our results confirm the findings of \citet{Svensmark06} and conjecture of 
\citet{Shaviv03} that the Archean Earth would have experienced a greatly reduced CR flux than is the case today.   If CR do play a significant role in cloud seeding, it seems plausible that the paleoclimate of early Earth could have been affected by reduced cloud formation. 

\subsubsection{The Termination Shock Distance}

While solar rotation plays the dominant role in secular CR modulation, the timescale for its evolution is hundreds of Myr.   \citet{Svensmark06b} further suggested that the CR variations would occur as the Sun passed through regions of enhanced star formation during its orbit through the Galaxy.  One additional CR modulation effect probed by our models is the variation in interstellar medium (ISM) pressure in which the solar system finds itself.   The heliopause contracts and expands according to the balance between the ISM pressure and the solar wind ram pressure.  The radius of the heliopause is then roughly proportional to the square root of the local ISM pressure.   

The ISM is largely in pressure equilibrium, with the pressure balance shared between gas thermal and dynamic pressures, magnetic fields and cosmic rays \citep[see, e.g., the reviews of][]{Ferriere01,Cox05}.  However, the gas density and temperature distribution is highly inhomogeneous, and, although these components are also in approximate pressure balance, significant variations in pressure are thought to exist.  Self-gravitating molecular clouds can have pressures several times that of the ambient medium.  \citet{Cox05} assesss the spiral arm and inter-arm region average pressure differences to be a factor of 2 and perhaps more.  In a study of the local bubble and environs, \citet{Lallement03} notes that the gas pressure in ``local fluff''  in which the Sun is currently located is between 5 and 10 times less than the pressure deduced from soft X-ray background measurements and that magnetic pressure making up the differences is ``hardly compatible with the observations''.  However, later work by \cite{Koutroumpa09} has shown that much of the soft X-ray background is from foreground heliospheric emission. therefore, the pressure of the local bubble surrounding the solar system is still highly uncertain.

The current picture of the ISM then indicates the Sun will pass through pressure differences of factors of perhaps up to ten.  For a relative velocity of the order of 25~km~s$^{-1}$---approximately the current solar velocity relative to the local ISM \citep{Lallement05}---and a scale size of significant pressure inhomogeneities of, say, 50~pc, the timescale for CF flux variations would be of the order of a few Myr.

A factor of two in pressure corresponds to a change in TS radius of 40\%, and our models indicate a CR flux change of a factor of 2 or so.  A factor of ten pressure change would lead to CR flux changes of more than an order of magnitude.  CR fluxes would be expected to be higher during spiral arm passages, where the average pressure is higher, but could also spike when passing through molecular clouds whose higher pressures could greatly compress the heliosphere.   The mechanism of heliospheric compression would augment (and perhaps dominate at times) the spiral arm passage increase in the ambient CR flux that \citet{Shaviv05} and \citet{Svensmark06} have suggested could have lead to $\sim250$~Myr cycles in the climate and biodiversity. \cite{Muller06} have studied the effect of the change in ISM pressure on the heliospheric size and structure using an MHD model, which includes the effect of  neutral atoms. Based on their model cases for the heliosphere, they calculated the change in CR intensity at 1 AU and found that the flux can differ by an order of magnitude, depending on the density and ionization state of the ISM region in which the solar system finds itself.  Compression of the heliosphere due to higher ISM pressure generally leads to larger CR fluxes at Earth in their models, consistent with the direction of the effect modeled much more simply here.  \citep{Fields08}  found that the passage of the blast from a supernova explosion 10pc away would shrink the heliopause down to about 1~AU.  Such ``astrospheric collapse'' has been estimated to occur with a frequency of 1--10~Gyr$^{-1}$ due to passage through very dense interstellar clouds by \citep{SmithScalo09}.


\acknowledgments
The majority of this work was supported by SI Grand Challenges grant number 40510254HH0022.  We thank the Director of the Unlocking the Mysteries of the Universe Consortium, Christine Jones, for support and encouragement.  
JJD was supported by NASA contract NAS8-39073 to the {\it Chandra X-ray Center} and thanks the Director, H.~Tananbaum, for advice and support.  JK was supported by NASA Grant NNX08AQ14G.





\end{document}